\documentstyle[11pt,aaspp4]{article}

\def\lalpha{L$\alpha$}
\def\halpha{H$\alpha$}
\def\hbeta{H$\beta$}
\def\hgamma{H$\gamma$}
\def\hdelta{H$\delta$}
\def\hepsilon{H$\epsilon$}
\def\lam{$\lambda$}
\def\micron{\ifmmode {\mu \rm m} \else {$\mu \rm m$}\fi}

\def\zabs{z$_{abs}$}
\def\zem{z$_{em}$}
\def\hnot{\ifmmode {\rm H_0} \else H$_0$\fi}

\def\lsun{\ifmmode {\rm L_\odot} \else L$_\odot$\fi}
\def\msun{\ifmmode {\rm M_\odot} \else M$_\odot$\fi}
\def\msunyr{\ifmmode {\rm M_\odot~yr^{-1}}\else${\rm M_\odot~yr^{-1}}$\fi}

\def\cmmitwo{\ifmmode \rm cm^{-2} \else $\rm cm^{-2}$\fi}
\def\cmmithree{\ifmmode \rm cm^{-3} \else $\rm cm^{-3}$\fi}
\def\cmps{\ifmmode \rm cm~s^{-1}\else $\rm cm~s^{-1}$\fi}
\def\cmpsps{\ifmmode \rm cm~s^{-2}\else $\rm cm~s^{-2}$\fi}
\def\kmps{\ifmmode \rm km~s^{-1}\else $\rm km~s^{-1}$\fi}
\def\kmpspmpc{\ifmmode \rm km~s^{-1}~Mpc^{-1} \else
    $\rm km~s^{-1}~Mpc^{-1}$\fi}
\def\ergps{\ifmmode \rm erg~s^{-1} \else $\rm erg~s^{-1}$ \fi}
\def\ergpsphz{\ifmmode \rm erg~s^{-1}~Hz^{-1} \else 
   $\rm erg~s^{-1}~Hz^{-1}$ \fi}
\def\mdotav{\ifmmode \langle \dot m_o \rangle \else 
   $\langle \dot m_o \rangle$ \fi}

\def\eg{e.g.}
\def\ie{i.e.}

\def\etal{~et al.}

\def\iras{{\it IRAS}}

\received{}
\accepted{}
\journalid{}{}
\articleid{}{}

\slugcomment{Submitted to P.A.S.P}

\begin{document}

\title{A Brief History of AGN}

\author{Gregory A. Shields}
\affil{Department of Astronomy, University of Texas, Austin, TX 78712}
\authoremail{shields@astro.as.utexas.edu}

\abstract

Astronomers knew early in the twentieth century that some galaxies
have emission-line nuclei.  However, even the systematic study
by Seyfert (1943) was not enough to launch active galactic
nuclei (AGN) as a major topic of astronomy.  The advances in
radio astronomy in the 1950s revealed a new universe of energetic
phenomena, and inevitably led to the discovery of quasars.
These discoveries demanded the attention of observers and
theorists, and AGN have been a subject of intense effort
ever since.
Only a year after the recognition of the redshifts of 3C 273
and 3C 48 in 1963, the idea of energy production by accretion
onto a black hole was advanced.  However, acceptance of this
idea came slowly, encouraged by the discovery of black hole
X-ray sources in our Galaxy and, more recently, supermassive
black holes in the center of the Milky Way and other galaxies.
Many questions remain as to the formation and fueling of the
hole, the geometry of the central regions, the detailed emission
mechanisms, the production of jets, and other aspects.
The study of AGN will remain a vigorous part of astronomy
for the foreseeable future.


\section{INTRODUCTION}

Although emission lines
in the nuclei of galaxies were recognized at the beginning 
of the twentieth century, a half century more would pass before 
active galactic nuclei (AGN) became a focus of intense research effort.
The leisurely pace of optical discoveries
in the first half of the century gave way to the fierce competition
of radio work in the 1950s.  The race has never let up.  Today,
AGN are a focus of observational effort in every frequency band 
from radio to gamma rays. Several of these bands 
involve emission lines as well as continuum.  AGN theory centers
on extreme gravity and black holes, among the most exotic concepts of
modern astrophysics. Ultrarelativistic particles, magnetic fields,
hydrodynamics, and radiative transfer all come into play. In
addition,  AGN relate to the
question of galactic evolution in general.  For most
of the time since the
recognition of quasar redshifts in 1963, these objects have reigned as
the most luminous and distant  objects in the Universe.
Their use as probes of intervening
matter on cosmic scales adds a further dimension to the
importance of AGN.

For all these reasons, the enormous effort to describe and explain AGN
in all their variety and complexity is quite natural.  
We are far from having a detailed and certain understanding of AGN.
However, the working hypothesis that they
involve at their core a supermassive black hole producing energy
by accretion of gas has little serious competition today.  If this picture
is confirmed, then the past decade may be seen as a time when
AGN research  shifted from guessing the nature of AGN
to trying to prove it.  

Although the story is not finished, this seems a good time
to take stock of the progress that has been made.  The
present short summary is intended to give students of
AGN an account of some of the key developments 
in AGN research.
The goal is to bring the story to the point 
where a contemporary review of
some aspect of AGN might begin its detailed discussion.  
Thus, various threads typically are followed
to a significant point in the 1980s.

I have attempted to trace
the important developments without excessive technical detail,
relying on published sources, my own recollections,
and conversations with a number of researchers. 
The focus is on the actual active nucleus.  Fascinating aspects such as
intervening absorption lines, statistical surveys, and links to galactic
evolution receive relatively little discussion.   The  volume of
literature is such that only a tiny fraction of the important papers can be
cited.

\section{BEGINNINGS}

Early in the twentieth century, Fath (1909) undertook at Lick Observatory
 a series of observations aimed at clarifying the
nature of the ``spiral nebulae''.  A major question at the time was
whether spirals were relatively nearby, gaseous
objects similar to the Orion nebula, or very distant collections
of unresolved stars.  Fath's goal was to test the claim that
spirals show a continuous spectrum consistent with a collection of
stars, rather than the bright line spectrum characteristic of
gaseous nebulae.  He constructed a spectrograph
designed to record the spectra of faint objects, mounted it
on the 36-inch Crossley reflector, and guided the long
exposures necessary to obtain photographic spectra of these 
objects. For most of his objects, Fath found a continuous spectrum
with stellar absorption lines, 
suggestive of an unresolved collection of solar type stars.  
However, in the
case of NGC 1068, he observed that the ``spectrum is composite,
showing both bright and absorption lines''.  The six bright lines
were recognizable as ones seen in the spectra of gaseous
nebulae.

The bright and dark lines of NGC 1068 were confirmed by 
Slipher (1917) with spectra taken in
 1913 at Lowell Observatory.  In 1917,
he obtained a spectrum with a narrow spectrograph slit, and found
that the emission lines were not images of the slit but rather
``small disks'', i.e., the emission was spread over a substantial
range of wavelengths.  (However,
 he rejected an
``ordinary radial velocity interpretation'' of the line widths.)
During the following years, several astronomers noted the
presence of nuclear emission lines in the spectra of
some spiral nebulae.  For example, Hubble (1926)
mentioned that the relatively rare spirals with stellar
nuclei show a planetary nebula type
spectrum, notably NGC 1068, 4051, and 4151. 

The systematic study
of galaxies with nuclear emission lines began with the
work of  Seyfert (1943).  Seyfert obtained spectrograms
of 6 galaxies with nearly stellar nuclei showing emission lines
superimposed on a normal G-type (solar-type) spectrum: NGC 1068, 1275, 3516,
4051, 4151, and 7469. 
The two brightest (NGC 1068, 4151) showed
``all the stronger emission lines ... in planetary nebulae like
NGC 7027.''  Seyfert attributed the large widths of the lines to
Doppler shifts, reaching up to 8,500 \kmps\ for the hydrogen lines
of NGC 3516 and 7469. The emission-line profiles differed from
line to line and from object to object, but two patterns
were to prove typical of this class of galaxy.  The
forbidden and permitted lines in NGC 1068 had roughly similar
profiles with widths of $\sim$3000 \kmps.  In contrast, NGC 4151
showed relatively narrow forbidden lines, and corresponding
narrow cores of the permitted lines; but the hydrogen lines
had very broad 
(7500 \kmps) wings that were absent from the profiles of the
forbidden lines.  Seyfert contrasted these spectra with the narrow
emission lines
of the diffuse nebulae (H II regions) seen in irregular
galaxies and in the arms of spiral galaxies.
Galaxies with high excitation nuclear emission lines are
now called ``Seyfert galaxies''.  However, Seyfert's paper
was not enough to launch the study of AGN as a major focus
of astronomers' efforts.  The impetus for this came from a
new direction -- the development of radio astronomy.  

Jansky (1932), working at the Bell Telephone Laboratories,
conducted a study of the sources of static affecting
trans-Atlantic radio communications.  Using a rotatable
antenna and a short-wave receiver operating at a wavelength of
14.6 m, he systematically measured the intensity of the
static arriving from all directions throughout the day.
From these records, he identified three types of static: (1) static
from local thunderstorms, (2) static from distant thunderstorms,
and (3) ``a steady hiss type static of unknown origin''.  The
latter seemed to be somehow associated with the sun (Jansky 1932).  Continuing
his measurements throughout the year, Jansky (1933) observed
that the source of the static moved around in azimuth every
24 hours, and the time and direction of maximum changed gradually
throughout the year in a manner consistent with the earth's
orbital motion around the sun.  He inferred that the radiation
was coming from the center of the Milky Way galaxy.  After
further study of the data, Jansky (1935) concluded that
the radiation came from the entire disk of the Milky Way, being
strongest in the direction of the Galactic center.

Few professional astronomers took serious note of Jansky's work,
and it fell 
to an engineer, working at home in his spare time, to advance
the subject of radio astronomy. Reber (1940a,b) built
a 31 foot reflector in his backyard
near Chicago.  
He published
a map of the radio sky at 160 MHz showing several local
maxima, including one in the constellation Cygnus
that would prove important for AGN studies (Reber 1944).  He also noted
that the ratio of radio radiation to optical light
was vastly larger for the Milky Way than the sun.

With the end of World War II, several groups of radio engineers
turned their efforts to the study of radio
astronomy.  Notable among these were the groups at Cambridge
and Manchester in England and at CSIRO
 in Australia.  The study
of discrete sources began with the accidental discovery of
a small, fluctuating source in Cygnus by Hey, Parsons, and Phillips
(1946) in the course of a survey of the Milky Way at 60 MHz.  With
their 6 degree beam, they set an upper limit of 2 degrees on the
angular diameter of the source.  The intensity fluctuations,
occurring on a time scale of seconds, were proved a few years
later to originate in the earth's ionosphere; but at first they served
to suggest that the radiation ``could only originate from
 a small number of discrete sources''.  The discrete nature of
the Cygnus source was confirmed by Bolton and Stanley (1948),
who used a sea-cliff interferometer to set an upper limit
of 8 arcmin to the width of the source.  These authors deduced a
brightness temperature of more than $4 \times 10^6$ K at 100 MHz and
concluded that a thermal origin of the noise was ``doubtful''.
Bolton (1948) published a catalog of 6 discrete sources
and introduced the nomenclature Cyg A, Cas A, etc.
 Ryle and Smith (1948) 
published results from
a radio interferometer at Cambridge  analogous to the optical interferometer
used by Michelson at Mt. Wilson to measure stellar diameters.
Observing at 80 MHz, they set an upper limit of 6 arcmin
to the angular diameter of the source in Cygnus.

Optical identifications of discrete sources (other than the sun)
were finally achieved by Bolton, Stanley, and
Slee (1949).
Aided by more accurate positions from sea cliff observations,
they identified Taurus A with the Crab Nebula supernova
remnant (M 1); Virgo A with M 87, a large elliptical galaxy
with an optical jet; and Centaurus A with NGC 5128,
an elliptical galaxy with a prominent dust lane.  
The partnership
of optical and radio astronomy was underway.

The early 1950s saw progress in radio surveys, position
determinations, and optical identifications.
A class of sources fairly uniformly distributed over the sky
was shown
by the survey by Ryle, Smith, and Elsmore (1950) based on observations
with the Cambridge interferometer.  
Smith (1951) obtained accurate positions
of four discrete sources, Tau A, Vir A, Cyg A, and Cas A.

Smith's positions enabled Baade and Minkowski (1954)
 to make optical identifications
of Cas A and Cyg A in 1951 and 1952. 
At the position of
Cyg A, they found an object with a distorted morphology, which they
proposed was two galaxies in
collision.  Baade and Minkowski found emission lines of
[Ne V], [O II], [Ne III], [O III], [O I], [N II], and \halpha,
with widths of about 400 \kmps.  The redshift of 16,830 \kmps\ 
implied a large distance, 31 Mpc, for the assumed
Hubble constant of $\hnot = 540~\kmpspmpc.$ 
The large distance
of Cyg A implied an enormous luminosity, $8 \times 10^{42}$ \ergps\
in the radio, larger than the optical
luminosity of $6 \times 10^{42}$ \ergps.  (Of course, these values
are larger for a modern
value of \hnot.)

This period also saw progress in the measurement of the structure
of radio sources.
Hanbury Brown, Jennison,
and Das Gupta (1952) reported results from the new 
intensity interferometer
developed at Jodrell Bank, including a demonstration that
Cyg A was elongated, with dimensions roughly 2 arcmin
by 0.5 arcmin.  
Interferometer measurements of Cyg A
by Jennison and Das Gupta (1952) showed two
equal components separated by 1.5 arcmin that straddled the
optical image, a puzzling morphology that proved to be common for
extragalactic radio sources.

Radio sources were categorized as
 `Class I' sources, associated
with the plane of the Milky Way, and `Class II' sources, isotropically
distributed and possibly mostly extragalactic (\eg, Hanbury Brown 1959).
Some of the latter had very small angular sizes, encouraging
the view that many were ``radio stars'' in our Galaxy.
Morris, Palmer, and Thompson (1957) published upper limits
of 12 arcsec on the size of 3 class II sources, implying
brightness temperatures in excess of $2 \times 10^7$ K.  They suggested
that these were extragalactic sources of the Cyg A type.

Theoretically, Whipple and Greenstein (1937) 
attempted to explain the Galactic radio background measured
by Jansky in terms of thermal emission by interstellar dust,
but the expected dust temperatures were far too low
to give the observed radio brightness.  Reber
(1940a) considered free-free emission by ionized gas in
the interstellar medium.  This process was considered more
accurately by Henyey and Keenan (1940) and Townes (1947), who
realized that Jansky's brightness temperature of $\sim 10^5~K$
could not be reconciled with thermal emission from interstellar
gas believed to have a temperature $\sim 10,000\ K$.  
Alfv\'en and Herlofson (1950) proposed
that ``radio stars'' involve cosmic ray electrons in a magnetic
field emitting by the synchrotron process.  
This quickly led Kiepenheuer (1950) to
explain the Galactic radio background in terms of synchrotron
emission by cosmic rays in the general Galactic magnetic field.
He showed order-of-magnitude agreement between the observed
and predicted intensities, supported by a more careful 
calculation by Ginzburg (1951).
The synchrotron explanation became accepted for extragalactic
discrete sources by the end of the 1950's.  The theory
indicated enormous energies, up to 
$\sim 10^{60}$ ergs for the ``double lobed'' radio galaxies
(Burbidge 1959).  The confinement of the plasma in these lobes
would later be attributed to ram pressure as the material 
tried to expand into the intergalactic medium
(De Young and Axford 1967).  A mechanism for production of
bipolar flows to power the lobes was given by the
``twin exhaust model'' of Blandford and Rees (1974).

The third Cambridge (3C) survey at 159 MHz 
(Edge \etal\ 1959) was  followed by the revised 3C survey
at 178 MHz (Bennett 1962).  Care was taken to
to minimize the confusion problems
of earlier surveys, and many radio sources
came to be known by their 3C numbers.  
These and the
surveys that soon followed provided many 
accurate radio positions 
as the search for
optical identifications accelerated.  
(AGN were also discovered in optical searches based
on morphological ``compactness'' [Zwicky 1964] and strong
ultraviolet continuum [Markarian 1967] and later infrared
and X-ray surveys.)  
Source counts as a function of 
flux density (``log N -- log S'') showed 
a steeper increase in numbers with decreasing flux density
than expected for a homogeneous, nonevolving universe with Euclidean
geometry (\eg, Mills, Slee, and Hill 1958; 
Scott and Ryle 1961).  This was used to argue against the
``steady state'' cosmology (Ryle and Clark 1961), 
although some disputed such
a conclusion (\eg, Hoyle and Narlikar 1961).

\section{THE DISCOVERY OF QUASARS}

Minkowski's
studies of radio galaxies culminated with
identification of 3C 295 with a member of a cluster
of galaxies at the unprecedented
redshift of 0.46 (Minkowski 1960).  
Allan Sandage of the Mt. Wilson
and Palomar Observatories and Maarten Schmidt of
the California Institute of Technology (Caltech) then took
up the quest for optical identifications and redshifts 
of radio galaxies.  Both worked with Thomas A. Matthews, who obtained
accurate radio positions with the new interferometer at the
Owens Valley Radio Observatory operated by Caltech. 
In 1960, Sandage obtained a
photograph of 3C 48 showing a $16^m$ stellar object with
a faint nebulosity.  The spectrum of
the object showed broad emission lines at unfamiliar wavelengths,
and photometry showed the object to be variable
and to have an excess of ultraviolet emission compared with
normal stars.  Several other apparently star-like images
coincident with radio sources were found to show strange,
broad emission lines.  Such objects
came to be known as quasi-stellar radio sources (QSRS),
quasi-stellar sources (QSS), or quasars.  Sandage reported
the work on 3C 48 in  an unscheduled paper in the December, 1960,
meeting of the AAS  (summarized by the editors of
{\it Sky and Telescope} [Matthews et al. 1961]).  There was a
``remote possibility that it may be a distant galaxy of stars''
but ``general agreement'' that it was ``a relatively nearby star
with most peculiar properties.''

The breakthrough came on February 5, 1963,
as Schmidt was pondering the spectrum
of the quasar 3C 273.  An accurate position had been obtained
in August, 1962 by Hazard, Mackey, and Shimmins (1963), who used the
210 foot antenna at the Parkes station in Australia to
observe a lunar occultation of 3C 273.  From the precise time
and manner in  which the source disappeared and reappeared,
they determined that the source had two components.
3C 273A had a fairly typical class II radio
spectrum, $F_{\nu} \sim
\nu^{-0.9}$; and it was separated by 20 seconds of arc from
component `B', which had a size less than 0.5 arcsec and
a ``most unusual'' spectrum, $f_{\nu} \sim \nu^{0.0}$.  
Radio positions B and A, respectively, 
coincided with those of a 13$^m$ star like
object and with a faint wisp or jet pointing away from
the star.  At first suspecting
the stellar object to be a foreground star, Schmidt 
obtained spectra of it at
the 200-inch telescope in late December, 1962.
The spectrum showed broad emission lines at unfamiliar wavelengths,
different from those of 3C 48.
Clearly, the object was no ordinary star. 
Schmidt noticed that four emission lines
in the optical spectrum showed a pattern of decreasing
strength and spacing toward the blue, reminiscent of the
Balmer series of hydrogen.  He found that the four lines
agreed with the expected wavelengths of \hbeta, \hgamma,
\hdelta, and \hepsilon\  with a redshift of z = 0.16.  This redshift
in turn allowed him to identify a line in the ultraviolet part
of the spectrum with Mg II \lam 2798.  
Schmidt consulted with his colleagues, Jesse L. Greenstein and
J. B. Oke.
Oke had obtained
photoelectric spectrophotometry of 3C 273 at the 100-inch telescope,
which revealed an emission-line in the infrared at \lam 7600.  With
the proposed redshift, this feature 
agreed with the expected wavelength of \halpha. 
Greenstein's spectrum of 3C 48 with a redshift of z = 0.37,
supported by the presence of Mg II in both objects.
The riddle of the spectrum of quasars was solved.

These results were published in {\it Nature} six weeks later
in adjoining papers
by Hazard et al. (1963); Schmidt (1963); Oke (1963); and Greenstein
and Matthews (1963).  The objects might be galactic stars with a very
high density, giving a large
gravitational redshift. However, this explanation was
difficult to reconcile with the widths of the
emission lines and the presence of forbidden
lines. The ``most direct and least objectionable''
explanation was that the objects were extragalactic,
with redshifts reflecting the Hubble expansion.  The redshifts 
were large but not unprecedented;
that of 3C 48 was second only to that of 3C 295.  
The radio
luminosities of the two quasars were comparable with those
of Cyg A and 3C 295.  However, the optical luminosities
were staggering, 
``10 - 30 times brighter than the brightest
giant ellipticals''; and the radio surface brightness
was larger than for the radio galaxies.  The redshift of 3C 273
implied a velocity of 47,400  \kmps\ and a distance of
about 500 Mpc (for $\hnot~\approx 100~\kmpspmpc$).  The nuclear
region would then be less than 1 kpc in diameter.  The jet would be
about 50 kpc away, implying a timescale greater than $10^5$ years
and a total energy radiated of at least $10^{59}$ ergs.

Before the redshift of 3C 273 was announced,
Matthews and Sandage (1963) had submitted a paper  
identifying 3C 48,
3C 196 and 3C 286 with stellar optical objects.
They explored the popular notion that these
objects were some kind of Galactic star, arguing
from their isotropic distribution on the sky and
lack of observed proper motion that the most
likely distance from the sun was about 100 pc.
The objects had peculiar colors, and 3C 48 showed
light variations of 0.4 mag.  In a section added following the
discovery of the redshifts of
3C 273 and 3C 48, they  pointed out that the size limit of $\le$0.15
pc implied by the optical light variations was important in the
context of the huge distance and luminosity implied by 
taking the redshift to result from the Hubble expansion.

A  detailed analysis of 3C48 and 3C 273 was published
by Greenstein and Schmidt (1964).  They considered
explanations of the redshift involving (1) rapid motion
of objects in or near the Milky Way, (2) gravitational
redshifts, and (3) cosmological redshifts. If 3C 273
had a transverse velocity comparable with the radial
velocity implied by its redshift, the lack of an observed
proper motion implied a distance of at least 10 Mpc
(well beyond the nearest galaxies).
The corresponding absolute magnitude was 
closer to the luminosity of galaxies than stars. 
The four quasars with known velocities were all receding;
and accelerating a massive, luminous
object to an appreciable fraction of the speed of light
seemed difficult.  Regarding gravitational redshifts,
Greenstein and Schmidt argued that the widths of the
emission lines required the line emitting gas to be
confined to a small fractional radius around the massive
object producing the redshift.  The observed symmetry
 of the line profiles seemed unnatural in a gravitational
redshift model.  For a 1~\msun\ object, 
the observed \hbeta\ flux
 implied an electron density
$N_e \approx 10^{19}$ cm$^{-3}$, incompatible with the observed
presence of forbidden lines in the spectrum. The emission-line
constraint, together with a requirement that the massive
object not disturb stellar orbits in the Galaxy, required
a mass $\ge 10^9$ \msun.  The stability of such a 
``supermassive star'' seemed doubtful in the light of theoretical
work by Hoyle and Fowler (1963a), who had examined such objects
as possible sources for the energy requirements of extragalactic
radio sources.  Adopting the cosmological explanation of the
redshift, Greenstein and Schmidt derived radii for
a uniform spherical emission-line region of 11 and 1.2 pc for 3C 48 and
3C 273, respectively.  This was based on the \hbeta\ luminosities
and electron densities estimated from the \hbeta, [O II], and [O
III] line ratios.  Invoking light travel time constraints based on
the observed optical variability (Matthews and Sandage 1963; Smith
and Hoffleit 1963),
 they proposed a model
in which a central source of optical continuum was surrounded by
the emission-line region, and a still larger radio emitting region.
They suggested that a central mass of order $10^9$ \msun\
might provide adequate energy for the  lifetime of $\ge 10^6$ yr
implied by the jet of 3C 273 and the nebulosity of 3C 48.
This mass was about right to confine the line emitting gas,
which would disperse quickly if it expanded at the observed
speeds of 1000 \kmps\ or more.  Noting that such a mass would
correspond to a Schwarzschild radius of $\sim 10^{-4}$ pc,
they observed that ``It would be important to know whether
continued energy and mass input from such a `collapsed'
region are possible''.  Finally, they noted that there
could be galaxies  around 3C 48 and 3C 273 
hidden by the glare of the nucleus.  Many features of this
analysis are recognizable in current thinking about AGN.

The third and fourth quasar redshifts were published by
Schmidt and Matthews (1964), who found 
z = 0.425 and  0.545 for 
3C 47 and 3C 147, respectively.
Schmidt (1965) published redshifts for 5 more quasars.
For 3C 254, a redshift z = 0.734, based
on several familiar lines, allowed the identification
of C III] \lam 1909 for the first time.  This in turn allowed 
the determination of redshifts of 1.029 and 1.037 from
\lam 1909 and \lam 2798 in 3C 245 and CTA 102,
respectively. (CTA is a radio source list from
the Caltech radio observatory.)  For
3C 287, a redshift of 1.055 was found from \lam 1909,
\lam 2798, and another first, C IV \lam 1550.
Finally, a dramatically higher redshift of 2.012 was
determined for 3C 9 on the basis of \lam 1550 and the
first detection of the Lyman $\alpha$ line 
of hydrogen at \lam 1215.
The redshifts were large enough
that the absolute luminosities depended significantly
on the cosmological model used.

Sandage (1965) reported the discovery of a large population
of radio quiet objects that otherwise appeared to resemble quasars. 
Matthews and Sandage (1963) had found that quasars
showed an ``ultraviolet excess'' when compared with 
normal stars on a color-color (U-B, B-V)
diagram.  This led to a search technique in which 
exposures in U and B were recorded on the same photographic
plate, with a slight positional offset, allowing rapid
identification of objects with strong ultraviolet continua.
Sandage noticed a number of such objects that did not
coincide with known radio sources.  These he called ``interlopers'',
``blue stellar objects'' (BSO),
or ``quasi-stellar galaxies'' (QSG).
\footnote[1]{  
Here we adopt the now common practice of using the
term ``quasi-stellar object'' (QSO) to refer to  
these objects regardless of radio luminosity
(Burbidge and Burbidge 1967).}  
Sandage found that at magnitudes fainter than 15, 
the UV excess objects
populated the region occupied by quasars on the color-color
diagram, whereas brighter objects typically 
had the colors of main sequence
stars.  The number counts of the BSOs as a function of apparent
magnitude also showed a change of slope at $\sim 15^m$,
consistent with an extragalactic population of objects at
large redshift.  Spectra showed that many of these objects
indeed had spectra with large redshifts, including
z = 1.241 for BSO 1.  Sandage estimated that
the QSGs outnumbered the radio loud quasars by a factor $\sim 500$,
but this was reduced by later work (\eg, Kinman 1965;
Lynds and Villere 1965).

The large redshifts of QSOs immediately made them potential tools
for the study of cosmological questions.
The rough similarity of the emission-line strengths of QSOs to
those observed, or theoretically predicted, for planetary nebulae
suggested that the chemical abundances were
roughly similar to those in our Galaxy (Sklovskii 1964; Osterbrock
and Parker 1966).  Thus these objects, suspected by many astronomers
to lie in the nuclei of distant galaxies, had reached fairly
``normal'' chemical compositions when the Universe was considerably
younger than today.

 The cosmological importance of redshifts high enough
to make
\lalpha\  visible was quickly recognized.  Hydrogen gas in intergalactic
space would remove light from the quasar's spectrum at
the local cosmological redshift, and continuously distributed
gas would erase a wide band of continuum to the short wavelength
side of the \lalpha\ emission line (Gunn and
Peterson 1965; Scheuer 1965).  Gunn and Peterson set a tight
upper limit to the amount of neutral hydrogen in intergalactic
space, far less than the amount that would significantly retard the
expansion of the Universe.

The study of discrete absorption features in quasar spectra
also began to develop.  An unidentified sharp line was observed
in the spectrum of 3C 48 by Greenstein and Schmidt (1964).  Sandage
(1965) found that the \lam 1550 emission line of BSO 1 was ``bisected
by a sharp absorption feature''.
The first quasar found with a rich absorption spectrum was 3C 191
(Burbidge, Lynds, and Burbidge 1966; Stockton and Lynds 1966).
More than a dozen sharp lines were identified, including \lalpha\
and lines of C II, III, and IV and Si II, III, and IV.
A rich set of narrow absorption lines was also observed
in the spectrum of PKS 0237-23, whose emission-line
redshift, z = 2.223, set a record at the time. Arp, Bolton, and
Kinman (1967) and Burbidge (1967a) respectively proposed
absorption line redshifts of z = 2.20 and 1.95 for this object, but
each value left many lines without satisfactory identifications.
It turned out that
both redshifts were present (Greenstein and Schmidt 1967).

All these absorption systems had
\zabs $<$ \zem.  They could be interpreted
as intervening clouds imposing absorption spectra at the appropriate
cosmological redshift, as had been anticipated theoretically
(Bahcall and Salpeter 1965).  Alternatively,
they might represent material expelled from the quasar,
whose outflow velocity is subtracted from the cosmological
velocity of the QSO.  However, PKS 0119-04 was found to have
\zabs $>$ \zem, implying material that was in some sense falling
into the QSO from the near side with a relative velocity of 10$^3$
\kmps\ (Kinman and Burbidge 1967).  Today, a large fraction
of the narrow absorption lines with \zabs\ substantially
less than \zem\ are believed to result from intervening
material.  This includes the so-called ``Lyman alpha forest''
of closely spaced, narrow \lalpha\ lines that punctuate the
continuum to the short wavelength side of the \lalpha\ emission line,
especially in high redshift QSOs.  The study of intervening
galaxies and gas clouds by means of absorption lines in the
spectra of background QSOs is now a major branch of astrophysics.

A different kind of absorption
was discovered in the spectrum of PHL 5200 by  Lynds (1967).
This object showed broad absorption bands on the short wavelength
sides of the \lalpha, N V \lam 1240, and C IV \lam 1550 emission
lines, with a sharp boundary between the emission and absorption.
Lynds interpreted this in terms of an expanding shell of gas around
the central object.  Seen in about 10 percent of radio quiet QSOs (Weymann
\etal\ 1991), these broad absorption lines (BALs) are among the many
dramatic but poorly understood aspects of AGN.

The huge luminosity of QSOs, rapid
variability, and implied small size caused some astronomers to
question the cosmological nature of the redshifts.
Terrell (1964) considered the possibility that the objects were
ejected from the center of our galaxy.  Upper limits on the proper
motion of 3C 273, together with a Doppler interpretation of the
redshift, then implied a distance of at least 0.3 Mpc and an age at
least 5 million years.  Arp (1966), pointing to close pairs of
peculiar galaxies and QSOs on the sky, argued for noncosmological
redshifts that might result from ejection from the peculiar galaxies
at high speeds or an unknown cause. Setti and Woltjer (1966) noted
that ejection from the Galactic center would imply for the QSO
population an explosion with energy at least $10^{60}$ ergs, and
more if ejected from nearby radio galaxies such as Cen A
as suggested by Hoyle and Burbidge (1966). 
Furthermore, Doppler boosting would cause us to see more blueshifts
than redshifts if the objects were ejected from nearby galaxies
(Faulkner, Gunn, and Peterson 1966).  Further evidence for
cosmological redshifts was provided by Gunn (1971), who showed
that two clusters of galaxies containing QSOs had the same
redshifts as the QSOs. Also, Kristian (1973) showed that the
``fuzz'' surrounding the quasistellar image of a sample of QSOs was
consistent with the presence of a host galaxy.

\section{CHARTING THE TERRAIN}

At this stage, a number of properties of AGN were recognized.  
Most astronomers
accepted the cosmological redshift of QSOs, 
and the parallel between Seyfert galaxies and
QSOs suggested a common physical phenomenon.
Questions included the nature of the energy source, 
the nature of the continuum source and  emission-line regions,
and the factors that produce an AGN in some galaxies and not others.

\subsection{Emission Lines}

The basic parameters of the region of gas emitting the narrow emission
lines were fairly quickly established.
In one of the first physical analyses of ``emission nuclei'' in galaxies,
Woltjer (1959) derived a density $\rm N_e \approx 10^4~\cmmithree$ 
and temperature $T \approx 20,000$~K from
the [S II] and [O III] line ratios of Seyfert galaxies.  
The region emitting the narrow lines was just resolved 
for the nearest Seyfert galaxies,
giving a diameter of order 100 pc (e.g., Walker 1968; Oke and Sargent 1968).  
Oke and Sargent derived a mass of $\sim 10^5~\msun$ 
and a small volume filling factor 
for the narrow line gas in NGC 4151. 
Burbidge, Burbidge, and Prendergast (1958) found 
that the nuclear emission lines
of NGC 1068 were much broader than could be accounted for by the rotation
curve of the galaxy, and concluded that the material was in a state
of expansion.

A key question was why, in objects showing broad wings,
these were seen on the
permitted lines but not the forbidden lines.   
(Seyfert galaxies
with broad wings
came to be called ``Seyfert 1'' or ``Sy 1'' and those without them ``Sy 2''
[Khachikian and Weedman 1974].)
Were these
wings emitted by the same gas that emits the narrow lines?  Woltjer (1959) 
postulated a separate region of fast moving, possibly gravitationally bound
gas to produce the broad Balmer line wings of Seyfert galaxies. 
 Souffrin (1969a) adopted such a model in her analysis of NGC 3516 and NGC
4151. Alternatively, broad Balmer line wings might
be produced by electron scattering (Burbidge \etal\
1966).  Oke and Sargent (1968) supported this possibility for NGC 4151.
Their analysis of the emission-line region gave an electron scattering optical
depth $\tau_e \sim 0.1$.  Multiple scattering of Balmer line photons 
by the line opacity might increase the effective 
electron scattering probability,
explaining the presence of wings only on the permitted lines.
However, analysis of
electron scattering profiles by other authors (e.g.,
Weymann 1970) indicated the need for a dense region only a tiny fraction
of a light year across.  Favoring mass motions were
the irregular broad
line profiles in some objects (Anderson 1971),
which demonstrated the presence of bulk velocities of the needed
magnitude.  
In addition, Shklovskii (1964) had argued for an electron scattering optical
depth $\tau_{es} < 1$ in 3C 273 to avoid excessive smoothing of the continuum
light variations.  The picture of broad lines from
a small region of dense,
fast moving clouds (``Broad Line Region'' or BLR) and narrow lines
from a larger region of slower moving, less dense clouds (``Narrow
Line Region'' or NLR) found support from photoionization
modes (Shields 1974).

Early workers (e.g.,
Seyfert 1943) had noted that the narrow line intensities 
resembled those of planetary nebulae,
and photoionization was an obvious candidate 
for the energy input to the emitting gas
for both the broad and narrow lines.  
For 3C 273, Shklovskii (1964) noted that the kinetic energy of the emission-
line gas could power the line emission only for a very short time, whereas
the extrapolated power in ionizing ultraviolet radiation was in rough
agreement with the emission line luminosities. Osterbrock and Parker
(1965) argued against photoionization because of the observed weakness of the
Bowen O III fluorescence lines.  Also eliminating  thermal collisional
ionization because of the observed wide range of ionization stages,  they
proposed ionization and heating by fast protons resulting from high velocity
cloud collisions.  Souffrin (1969b) rejected this on the basis of 
thermal equilibrium considerations,
and argued along with Williams and Weymann (1968) 
that thermal collisional ionization was inconsistent with
observed temperatures.
Noting that an optical-ultraviolet continuum of roughly the needed power is
observed, and that the thermal equilibrium gives roughly the observed
temperature,  Souffrin concluded that
a nonthermal ultraviolet continuum  was 
``the only important source of ionization''.
Searle and Sargent (1968) likewise noted that the 
equivalent widths of the broad 
\hbeta\ emission lines were similar among AGN over a wide range of luminosity 
and were consistent with an extrapolation of the observed ``nonthermal''
continuum as a power law to ionizing frequencies. 
Detailed models of gas clouds photoionized by a power-law continuum were
calculated with the aid of electronic computers, with application to the Crab
nebula, binary X-ray sources, and AGN (Williams 1967; Tarter and
Salpeter 1969; Davidson 1972; MacAlpine 1972).
Such models showed that photoionization can account
for the intensities of the strongest optical 
and ultraviolet emission lines.  
In particular, the penetrating high frequency photons can
explain the simultaneous presence of 
very high ionization stages and strong emission
from low ionization stages, in the context of a ``nebula'' that is optically
thick to the ionizing continuum.  Photoionization quickly became accepted as
the main source of heating and ionization in the emission-line gas. 

Attention then focussed on improving photoionization models and understanding
the geometry and dynamics of the gas emitting the broad lines.  It was clear
that the emitting gas had only a tiny volume filling factor, and  one possible
possible geometry was the traditional nebular picture of clouds or
``filaments'' scattered through the BLR volume. 
Photoionization models typically assumed a slab geometry representing the
ionized face of a cloud that was optically thick to the Lyman continuum. 
Model parameters included the density and chemical composition of the gas and
the intensity and energy distribution of the incident ionizing continuum.
Various line ratios, such as C III]/C IV, were used to constrain the
``ionization parameter'', \ie, the ratio of ionizing photon density to gas
density.  Chemical abundances were assumed to be
approximately solar but were hard to determine because the high densities
prevented a direct measurement of the electron temperature from available
line ratios.

A challenge for photoionization models was the discovery that the 
\lalpha/\halpha\ ratio was an order-of-magnitude smaller than the value $\sim
50$ predicted by photoionization models at the time  (Baldwin 1977a; Davidsen,
Hartig, and Fastie 1977).  This stimulated models
with an improved treatment of radiative transfer in optically thick hydrogen
lines (\eg, Kwan and Krolik 1979).  These models found strong Balmer line
emission from a ``partially ionized zone'' deep in the cloud, heated by
penetrating X-rays, from which Lyman line emission was unable to escape.
The models still did not do a perfect job of explaining the observed
ratios (\eg, Lacy \etal\
1982) of the Paschen, Balmer, and Lyman lines.  Models by
Collin-Souffrin, Dumont, and Tully (1982) and Wills, Netzer, and Wills (1985)
suggested the need for densities as high as $N_e \approx 10^{11}~\cmmithree$ to
explain the \halpha/\hbeta\ ratio.

The X-ray heated region also was important for the formation of the strong Fe
II multiplet blends observed in the optical and ultraviolet.  Theoretical
efforts by several authors culminated in models involving thousands of Fe
lines, with allowance for the fluorescent interlocking of different lines
(Wills \etal\ 1985).  These models enjoyed some success in
explaining the relative line intensities, but the total energy in the Fe II
emission was less than observed.  Although some of this discrepancy might
involve the iron abundance, Collin-Souffrin \etal\ (1980) proposed a
separate Fe II emitting region with a high density ($N_e \approx
10^{11}~\cmmithree$) heated by some means other than photoionization.  This
region might be associated with an accretion disk.  The Fe II emission and
the Balmer continuum emission that combined to form the 3000~\AA\ ``little
bump'' still are not fully explained, nor is the tendency for radio loud AGN
to have weaker Fe II and steeper Balmer decrements than radio quiet objects
(Osterbrock 1977).

A tendency for the equivalent width of the C IV emission line to decrease
with increasing luminosity was found by Baldwin (1977b).  Explanations of
this involved a possible decrease, with increasing luminosity, in the
ionization parameter and in the ``covering factor'', \ie, the fraction
($\Omega/4\pi$) of the ionizing continuum intercepted by the BLR gas
(Mushotzky and Ferland 1984).  The ionization parameter was also the
leading candidate to explain the difference in ionization level between
classical Seyfert galaxies and the ``low ionization nuclear emission regions''
or ``LINERs'' (Heckman 1980; Ferland and Netzer 1983; Halpern and Steiner
1983).

 The geometry and state of motion of the BLR
gas has been a surprisingly stubborn problem.  If the BLR was a swarm of
clouds, they might be falling in (possibly related to the accretion supply),
orbiting, or flying out.  Alternatively, the gas might be associated with an
accretion disk irradiated by the ionizing continuum (\eg, Shields 1977;
Collin-Souffrin 1987).  Except for the BAL QSOs,
there was little evidence for blueshifted absorption analogous to the P Cygni
type line profiles of stars undergoing vigorous mass loss.  The approximate
symmetry of optically thick lines such as \lalpha\ and
\halpha\ suggested that the motion was circular or random rather than
predominantly radial (\eg,  Ferland, Netzer, and Shields 1979).  However,
for orbiting (or infalling) gas, the line widths implied rather large masses
for the central object, given prevailing estimates of the BLR radius. In
addition, gas in Keplerian orbit seemed likely to give a double peaked line
profile or to have other problems (Shields 1978a).  In the face of these
conflicting indications, the most common assumption was that the gas took the
form of clouds flying outward from the central object.  The individual clouds
would disperse quickly unless confined by some intercloud medium, and a
possible physical model was provided by the two-phase medium discussed by
Krolik, McKee, and Tarter (1981).  Radiation pressure of the ionizing
continuum, acting on the bound-free opacity of the gas, seemed capable of
producing the observed velocities and giving a natural explanation of the
``logarithmic'' shape of the observed line profiles (Mathews 1974;
Blumenthal and Mathews 1975).  Interpretation of the line profiles was
complicated by the recognition of systematic offsets in velocity between the
high and low ionization lines (Gaskell 1982; Wilkes and Carswell 1982; Wilkes
1984)

A powerful new tool was provided by the use of
``echo mapping'' or ``reverberation mapping'' of the BLR.  Echo
mapping relies on the time delays between the continuum and line variations
caused by the light travel time across the BLR (Blandford and McKee
1982).   Early results showed that the BLR is smaller and denser than most
photoionization models had indicated (Ulrich \etal\ 1984; Peterson \etal\
1985).  Masses of the central object,
by this time assumed to be a black hole, could be derived with increased
confidence.  The smaller radii implied smaller masses that seemed
reasonable in the light of other considerations, and the idea of
gravitational motions for the BLR gained in popularity.  This was supported
by the rough tendency of the line profiles to vary symmetrically, 
consistent with ``chaotic'' or circular motions (\eg, Ulrich
\etal\ 1984).

\subsection{Energy Source}

The question of the ultimate energy source for AGN stimulated creativity even
before the discovery of QSO redshifts.  The early concept of radio galaxies
as galaxies in collision gave way to the recognition of galactic nuclei as
the sites of concentrated, violent activity.  Burbidge (1961) suggested
that a chain reaction of supernovae (SN) could occur in a dense star cluster
in a galactic nucleus.  Shock waves from one SN would compress neighboring
stars, triggering them to explode in turn.  Cameron (1962) considered
a coeval star cluster leading to a rapid succession of SN as the massive stars
finished their short lives.  Spitzer and Saslaw (1966),
building on earlier suggestions, developed another model involving a dense star
cluster.  The cluster core would evolve to higher star densities through
gravitational ``evaporation'', and this would lead to frequent stellar
collisions and tidal encounters, liberating large amounts of gas.
Additional ideas involving dense star clusters included pulsar
swarms (Arons, Kulsrud, and Ostriker 1975) and starburst models
(Terlevich and Melnick 1985).

Hoyle and Fowler (1963a,b) discussed the idea of a supermassive
star (up to $\sim10^8~\msun$) as a source of gravitational and thermonuclear
energy.  In additional to producing large amounts of energy per unit mass,
all these models seemed capable of accelerating particles to relativistic
energies and producing gas clouds ejected at speeds of
$\sim 5000~\kmps$, suggestive of the broad emission-line wings of Seyfert
galaxies.  In this
regard, Hoyle and Fowler (1963a) suggested that ``a magnetic
field could be wound toroidally between the central star and a surrounding
disk.''  The field could store a large amount of energy, leading to powerful
``explosions'' and jets like that of M87. Hoyle and Fowler (1963b) suggested
that ``only through the contraction of a mass of $10^7 - 10^8~\msun$ to the
relativistic limit can the energies of the strongest sources be obtained.''

Soon after, Salpeter (1964)  and 
Zeldovich (1964) proposed the idea of QSO energy production from
accretion onto a supermassive black hole.  For material gradually spiraling to
the innermost stable orbit of a nonrotating black hole at $r =
6GM/c^2$, the energy released per unit mass would be $0.057c^2$, enough to
provide the energy of a luminous QSO from a reasonable mass.  Salpeter imagined
some kind of turbulent transport of angular momentum, allowing 
the matter to move
closer to the hole, which would grow in mass during the accretion process.  

The black hole model received limited attention until Lynden-Bell (1969)
argued that dead quasars in the form of ``collapsed bodies'' 
(black holes) should
be common in galactic nuclei, given the lifetime energy output of quasars and
their prevalence at earlier times in the history of the universe.  
Quiescent ones
might be detectable through their effect on the mass-to-light ratio of nearby
galactic nuclei.  Lynden-Bell explored the thermal radiation and fast particle
emission to be expected in a disk of gas orbiting the hole, with energy
dissipation related to magnetic and turbulent processes.  For QSO
luminosities, the disk would have a maximum effective temperature of $\sim
10^5~\rm K$, possibly leading to photoionization and broad line emission.  He
remarked that ``with different values of the [black hole mass and accretion
rate] these disks are capable of providing an explanation for a large
fraction of the incredible phenomena of high energy astrophysics, including
galactic nuclei, Seyfert galaxies, quasars and cosmic rays."

Further evidence for relativistic conditions in AGN came from other
theoretical arguments. Hoyle,
Burbidge, and Sargent (1966)  noted that  relativistic electrons emitting 
optical and infrared
synchrotron radiation would also Compton scatter ambient photons, 
boosting their
energy by large factors. This would lead to ``repeated stepping
up of the energies of quanta'', yielding a divergence that came to be known
as the ``inverse Compton catastrophe''.  This would be attended by
rapid quenching of the energy of the electrons.  They argued that 
this supported the
idea of noncosmological redshifts.  In response, Woltjer (1966) invoked a model
with electrons streaming radially on field lines, which could greatly reduce
Compton losses.  He further noted that because ``the relativistic electrons and
the photons they emit both move nearly parallel to the line of sight, the time
scale of variations in emission can be much shorter than the size of the region
divided by the speed of light.''  The emission would also likely be anisotropic,
reducing the energy requirements for individual objects.  

\subsection{Superluminal Motion}

Dramatic confirmation of the suspected relativistic motions came from
the advancing technology of radio astronomy. 
Radio astronomers using conventional interferometers had shown that
many sources had structure on a sub-arcsec scale. 
Scintillation of the radio signal from some AGN, caused by the
interplanetary medium of our solar system, also implied sub-arcsec dimensions
(Hewish, Scott, and Wills 1964).  The compact radio sources in
some AGN showed flat spectrum components and  
variability on  timescales of months (Dent 1965; Sholomitsky 1965). 
The variability  suggested milliarcsec dimensions on the basis of light
travel time arguments.   The spectral shape and evolution found explanation
in terms of multiple, expanding components that were optically thick to
synchrotron self-sbsorption, which causes a low frequency cutoff in the
emitted continuum (Pauliny-Toth and Kellermann 1966, and
references therein).  Such models had interesting theoretical consequences,
including angular sizes (for cosmological redshifts) as small as
$10^{-3}$ arcsec, and large amounts of energy in relativistic electrons, far
exceeding the energy in the magnetic field.

These inferences made clear the need for angular resolution finer than was
practical with conventional radio interferometers connected by wires or
microwave links.  This was achieved by recording
the signal from the two antennas separately on magnetic tape, and correlating
the recorded signals later by analog or digital means. This technique came to
be known as ``very long baseline interferometry'' (VLB, later VLBI).
 After initial difficulties finding ``fringes'' in
the correlated signal, competing groups in Canada and the United States
succeeded in  observing several AGN in the spring of 1967, over baselines of
roughly 200~km (see Cohen \etal\
1968). The U.S. experiments typically used the 140 foot antenna at the
National Radio Astronomy Observatory in Green Bank, West Virginia, in
combination with increasingly remote
 antennas in Maryland, Puerto Rico,
 Massachusetts, California, and Sweden.  The latter gave an angular resolution
of 0.0006 arcsec. Within another year, observations were made between Owens
Valley, California, and Parkes, Australia, a baseline exceeding 10,000 km or
80 percent of the earth's diameter.  A number of AGN showed components
unresolved on a scale of $10^{-3}$ arcsec.

On October 14 and 15, 1970,  Knight \etal\ (1971) observed  quasars at
7840 MHz with the Goldstone, California - Haystack, Massachusetts
``Goldstack'' baseline.  3C 279 showed fringes consistent with a symmetrical
double source separated by
$(1.55 \pm 0.03) \times 10^{-3}$ arcsec.  Later observations on February 14
and 26, 1971, by Whitney \etal\ (1971) showed a double source structure at
the same position angle, but separated by a 
distinctly larger angle of $(1.69 \pm 0.02)
\times 10^{-3}$ arcsec.  Given the distance implied by the redshift of 0.538,
this rate of angular separation corresponded to a linear separation rate of
ten times the speed of light!  Cohen \etal\ (1971), also using Goldstack data,
observed ``superlight expansion'' in 3C 273 and 3C 279.  Whitney
\etal\ and Cohen \etal\ considered a number of interpretations of their
observations, including multiple components that blink on and off (the
``Christmas tree model'') and noncosmological redshifts.  However,  most
astronomers quickly leaned toward an explanation involving motion of emitting
clouds ejected from the central object at speeds close to, but not exceeding,
the speed of light.  
Rees (1966) had
calculated the appearance of relativistically expanding sources, and
apparent expansion speeds faster than that of light were predicted.   A picture
emerged in which a stationary component was associated with the central object,
and clouds were ejected at intervals of several years along a fairly stable
axis.  (Repeat ejections were observed in the course of time by VLBI
experiments.)  If this ejection occurred in both
directions, it could  supply energy to the extended double
sources.  The receding components would be greatly 
dimmed by special relativistic
effects, while the approaching components were brightened.  The two observed
components are then associated with the central object 
and the approaching cloud,
respectively. The fact that the two observed components had roughly equal
luminosities found an explanation in the relativistic jet model
of Blandford and K\"onigl (1979).

Apparent superluminal motion has now been seen in a number of quasars and
radio galaxies, and a possibly analogous phenomenon has been
observed in connection with black hole systems of stellar mass in our Galaxy
(Mirabel and Rodriguez 1994)

\subsection{X-rays from AGN}

One June 18, 1962, an Aerobee sounding rockets blasted skyward from White
Sands proving ground in New Mexico.  It carried a Geiger counter designed
 to detect astronomical sources of X-rays. 
 The experiment,
carried out by Giacconi \etal\ (1962), discovered
an X-ray background and a ``large peak'' in a 10 degree error box near the
Galactic center and the constellation Scorpius. A rocket experiment by Bowyer
\etal\ (1964) also found an isotropic background, confirmed the Scorpius source,
and detected X-rays from the Crab nebula.  Friedman
and Byram (1967) identified X-rays from the active galaxy M 87.   A
rocket
carrying collimated proportional counters sensitive in the 1 to 10 keV energy
range, found sources coincident with 3C 273, NGC 5128 (Cen A), and M87  (Bowyer,
Lampton, and Mack 1970).  The positional error box for 
3C 273 was small enough to
give a probability of less that
$10^{-3}$ of a chance coincidence.  The X-ray luminosity, quoted as $\sim
10^{46}~\ergps$,  was comparable with quasar's optical luminosity.

The first dedicated X-ray astronomy satellite, {\em Uhuru}, was launched in
1970.  Operating until 1973, it made X-ray work 
a major branch of astronomy.  X-rays were reported from the Seyfert galaxies
NGC 1275 and NGC 4151 (Gursky \etal\ 1971).
  The spectrum of NGC 5128 was consistent with a power law
of energy index $\alpha = -0.7$, where $\rm L_\nu \propto \nu^\alpha$; and
there was low energy absorption corresponding to  
 a column density of $9 \times 10^{22} ~\rm atoms~cm^{-2}$, 
possibly caused by gas
in the nucleus (Tucker
\etal\ 1973).
Early variability studies were hampered by the need to compare results from
different experiments, but Winkler and White (1975) found a large change
in the flux from Cen A in only 6 days from {\em OSO-7} data.  Using {\em Ariel
V} observations of NGC 4151, Ives \etal\ (1976) found a significant increase
in flux from earlier {\em Uhuru} measurements.  Marshall \etal\ (1981),
using Ariel V data on AGN gathered over a 5 year period, found that roughly
half of the sources varied by up to a factor of 2 on times less than or equal
to a year.  A number of sources varied in times of 0.5 to 5 days.  Marshall
\etal\ articulated the importance of X-ray variability observations, which
show that the X-rays ``arise deep in the nucleus'' and ``relate therefore to
the most fundamental aspect of active galaxies, the nature of the central
`power house'.''  

Strong X-ray emission as a characteristic of Sy 1 galaxies was established
by Martin Elvis and his coworkers from {\em Ariel V} data (Elvis \etal\ 1978).
This work increased to 15 the number of known Seyfert X-ray sources, of which
at least three were variable.  Typical luminosities were $\sim 10^{42.5}$ to
$10^{44.5}~\ergps$.  The X-ray power correlated with the infrared and optical
continuum and \halpha\ line.  Seyfert galaxies evidently made a significant
contribution to the X-ray background, and limits could be set on the
evolution of Seyfert galaxy number densities and X-ray luminosities in order
that they not exceed the observed background.  Elvis \etal\ considered thermal
bremsstrahlung ($10^7~\rm K$), synchrotron, and synchrotron self-Compton
models of the X-ray emission.

{\em HEAO-1}, the first of the {\em High Energy Astronomy Observatories},
was an X-ray facility that
operated from 1977 to 1979.   It gathered data on a sufficient sample of
objects to allow comparisons of different classes of AGN and to construct a
log N-log S diagram and improved luminosity function.  
{\em HEAO-1} provided broad-band X-ray spectral information for a
substantial set of AGN, showing spectral indices $\alpha \approx -0.7$,
with rather little scatter, and absorbing columns 
$<5\times10^{22}~\rm cm^{-2}$ (Mushotzky \etal\ 1980).

The {\em Einstein
Observatory (HEAO-2)} featured grazing incidence focusing optics allowing
detection of sources as faint as $\sim 10^{-7}$ the intensity of the Crab
nebula.  Tananbaum \etal\ (1979) used {\em Einstein} data to study QSOs as a
class of X-ray emitters.  Luminosities of $10^{43}$ to $10^{47}~\ergps$
(0.5 to 4.5 keV) were found.  OX169 varied substantially in under 10,000 s,
indicating a small source size.  This suggested a black hole mass not
greater than
$2 \times 10^8~\msun$, if the X-rays came from the inner portion of an
accretion flow.  By this time, strong X-ray emission was established as a
characteristic of all types of AGN and a valuable diagnostic of their
innermost workings.

\subsection{The Continuum}

Today, the word ``continuum'' in the context of AGN might bring to mind
anything from radio to gamma ray frequencies.  However, in the early
days of QSO studies, the term generally meant the optical continuum, extending
to the ultraviolet and infrared as observations in these bands became
available.  Techniques of photoelectric
photometry and spectrum scanning were becoming established as QSO studies
began.  The variability of QSOs, including 3C 48 and 3C 273 (e.g., Sandage
1963), was known and no doubt contributed to astronomers' initial hesitation
to interpret QSO spectra in terms of large redshifts.  In his contribution to
the four discovery papers on 3C 273, Oke (1963) presented spectrophotometry
showing a continuum slope $L_\nu \propto \nu^{+0.3}$ in the optical,
becoming redder toward the near infrared.  He noted that the energy
distribution did not resemble a black body, and inferred that there must be a
substantial contribution of synchrotron radiation.

A key issue for continuum studies has been the relative importance of thermal
and nonthermal emission processes in various wavebands.  Early work tended to
assume synchrotron radiation, or ``nonthermal emission'', in the absence of
strong evidence to the contrary.  The free-free and bound-free emission from
the gas producing the observed emission lines was generally a small
contribution.  The possibility of thermal emission from very hot gas was
considered for some objects such as the flat blue continuum of 3C 273 (e.g.,
Oke 1966).  The energy distributions tend
to slope up into the infrared; and for thermal emission from optically thin
gas, this would would have required a rather low temperature and an excessive
Balmer continuum jump.  This left the possibilities of nonthermal emission or
thermal emission from warm dust, presumably heated by the ultraviolet
continuum.

Observational indicators of thermal or nonthermal emission include
broad features in the energy distribution,
variability, and polarization.  For the infrared, one also has correlations
with reddening, the silicate
absorption and emission features, and  possible  angular
resolution of the source  (Edelson \etal\ 1988). For some objects, rapid
optical variability implied brightness temperatures that clearly required a
nonthermal emission mechanism.  For example, Oke (1967) observed day-to-day
changes of 0.25 and 0.1 mag for 3C 279 and 3C 446, respectively.  For many
objects, the energy distributions were roughly consistent with a power law of
slope near $\nu^{-1.2}$.  Power laws of similar slopes were familiar
from radio galaxies and the Crab nebula, where the emission extended through
the optical band. These spectra were interpreted in terms of synchrotron
radiation with  power-law energy distributions for the radiating,
relativistic electrons.  Such a power-law energy distribution was also
familiar from studies of cosmic rays, and thus power laws seemed natural in
the context of high energy phenomena like AGN. In addition to simple
synchrotron radiation, there might be a hybrid
process involving synchrotron emission in the submillimeter and far infrared,
with some of these photons boosted to the optical by ``inverse'' Compton
scattering (Shklovskii 1965).  The idea of a nonthermal continuum in the
optical, whose high frequency extrapolation provided the ionizing radiation
for the emission-line regions, was widely held for many years.  This was
invoked not only for QSOs but also for Seyfert galaxies, where techniques such
as polarization were used to separate the ``nonthermal" and galaxy components
(\eg, Visvanathan and Oke 1968).

Infrared observations were at first plagued by low sensitivity and inadequate
telescope apertures.
Measurements  of 3C 273 in the K filter (2.2 \micron), published by
Johnson (1964) and Low and Johnson (1965), showed a continuum steeply rising
into the infrared.
  Infrared
radiation from NGC 1068 was observed by  Pacholczyk and Wisniewski (1967),
also with a flux density ($F_\nu$) strongly rising to the longest wavelength
observed (``N'' band, or 10
\micron).  The infrared radiation dominated the power output of this object. 
  Becklin \etal\ (1973) found that much of the 10 \micron\
emission from NGC 1068 came from a resolved source 1 arcsec (90 pc) across and
concluded that most of the emission was not synchrotron emission.  In
contrast, variability of the 10 \micron\ emission from 3C 273 (\eg, Rieke
and Low 1972) pointed to a strong nonthermal component.  Radiation from hot
dust has a minimum source size implied by the black body limit on the surface
brightness, and this is more stringent for longer wavelengths radiated by
cooler dust. This in turn implies a minimum variability timescale as a
function of wavelength.  The near infrared emission of NGC 1068 was found to
be strongly polarized (Knacke and Capps 1974).

Improving infrared technology, and optical instruments such as the
multichannel spectrometer on the 200-inch telescope (Oke 1969), led
to larger and better surveys of the AGN continuum. 
Oke, Neugebauer, and Becklin (1970) reported observations of 28 QSOs from 0.3
to 2.2~\micron.  The energy distributions were similar in radio loud
and radio quiet QSOs.  They found that the energy
distributions could generally be described as a power law (index
-0.2 to -1.6 for $F_\nu\propto\nu^\alpha$) and that they remained ``sensibly
unchanged'' during the variations of highly variable objects.  Penston
\etal\ (1974) studied the continuum from 0.3 to 3.4
\micron\ in 11 bright Seyfert galaxies.  All turned up toward the
infrared, and consideration of the month-to-month 
variability pointed to different
sources for the infrared and optical continua.  From an extensive survey  of
Seyfert galaxies, Rieke (1978) concluded that strong infrared emission was a
``virtually universal'' feature,  and that the energy distributions in general
did not fit a simple power law.  The amounts of dust required were roughly
consistent with the expected dust in the emission-line gas of the active nucleus
and the surrounding interstellar medium.  A consensus emerged that the infrared
emission of Seyfert 2's was thermal dust emission, but the situation for Seyfert
1's was less clear (\eg, Neugebauer
\etal\ 1976, Stein and Weedman 1976).  From a  survey of the optical and
infrared energy distribution of QSOs, Neugebauer \etal\ (1979) concluded that
the slope was steeper in the 1-3
\micron\ band than in the 0.3-1 \micron\ band, and that an apparent broad
bump around 3 \micron\ might be dust emission.  Neugebauer \etal\ (1987)
obtained energy distributions from 0.3 to 2.2 \micron\ for the complete set
of quasars in the Palomar-Green (PG) survey (Green, Schmidt, and Liebert
1986) as well as some longer wavelength observations.  A majority of objects
could be fit with two power laws ($\alpha \approx -1.4$ at lower frequencies,
$\alpha \approx -0.2$ at higher frequencies) plus a ``3000 \AA\ bump''.

Measurements at shorter and
longer wavelengths were facilitated by the {\em International Ultraviolet
Explorer} (IUE) and the {\em Infrared Astronomical Satellite} (IRAS),
launched in 1978 and 1983, respectively.  Combining such measurements with
ground based data, Edelson and Malkan (1986) studied the spectral energy
distribution of AGN over the wavelength range 0.1-100
\micron.  The 3-5 \micron\ ``bump'' was present in most Seyferts and QSOs,
involving up to 40 percent of the luminosity between 2.5 and 10 \micron.  All
Sy 1 galaxies without large reddening appeared to require a hot thermal
component, identified with the increasingly popular concept of emission from
an accretion disk.  Edelson and Malkan (1987) used IRAS observations to study
the variability of AGN in the far infrared.  The high polarization objects
varied up to a factor 2 in a few months, but no variations greater than 15
percent were observed for ``normal'' quasars or Seyfert galaxies.  The former
group was consistent with a class of objects known as
``blazars'' that are dominated at all wavelengths by a variable, polarized
nonthermal continuum. 
Blazars were found to be highly variable at all wavelengths, but most AGN
appeared to be systematically less variable in the far infrared than at
higher frequencies.  This supported the idea of thermal emission from dust in
the infrared.  This was further supported by observations at submillimeter
wavelengths that showed a very steep decline in flux longward of the infrared
peak at around 100 \micron.  For example, an upper limit on the flux from NGC
4151 at 438
\micron\ (Edelson \etal\ 1988) was so far below the measured flux at 155
\micron\ as to require a slope steeper than
$\nu^{+2.5}$, the steepest that can be obtained from a self-absorbed
synchrotron source without special geometries.  Dust emission could explain a
steeper slope because of the decreasing efficiency of emission toward longer
wavelengths.

 Sanders \etal\ (1989)
presented measurements of 109 QSOs from 0.3 nm to 6 cm ($10^{10} - 10^{18}$
Hz).  The gross shape of the energy distributions was quite similar for most
objects, excepting the flat spectrum radio loud objects such as 3C 273.  This
typical energy distribution could be fit by
 a hot accretion disk at shorter
wavelengths and heated dust at longer wavelengths. 
Warping of the disk at larger radii was invoked to give the needed amount of
reprocessed radiation as a function of radius.  As noted by Rees
\etal\ (1969) and others, the rather steep slope in the infrared, giving rise
to an apparent minimum in the flux around 1 \micron, could be explained
naturally by the fact that grains evaporate if heated to temperatures above
about  1500~K. Sanders \etal\ saw ``no convincing evidence for
energetically significant nonthermal radiation'' in the wavelength range 3 nm
to 300 \micron\ in the continua of radio quiet and steep-spectrum 
radio-loud quasars. 
This paper marked the culmination of a gradual shift of sentiment
from nonthermal to thermal explanations for the continuum of non-blazar AGN.

The blazar family comprised ``BL Lac objects'' and ``Optically Violent
Variable'' (OVV) QSOs.  BL Lac objects, named after the prototype object
earlier listed in catalogs of variable stars, had a nonthermal continuum
but little or no line emission.  OVVs have the emission lines of QSOs.
These objects all show a continuum that is fairly well described as a power
law extending from X-ray to infrared frequencies.  They typically show rapid
(sometimes day-to-day) variability and strong, variable polarization.  The
continuum in blazars is largely attributed to nonthermal processes
(synchrotron emission and inverse Compton scattering).  3C 273
seems to be a borderline OVV  (Impey, Malkan, and Tapia 1989). The need for
relativistic motions, described above, arises in connection with this class
of objects.  A comprehensive study of the energy distributions of blazars
from $10^8$ to
$10^{18}$ Hz was given by Impey and Neugebauer (1988).  Bolometric
luminosities ranged from $10^9$ to $10^{14}~\lsun$, dominated by the 1 to 100
~\micron\ band. There was evidence for a thermal infrared component in many of
the less luminous objects, and an ultraviolet continuum bump 
associated with the presence of emission lines. When gamma rays are observed
from AGN (\eg, Swanenburg \etal\ 1978), they appear to be associated with the
beamed nonthermal continuum. The relationship of blazars to ``normal'' AGN is
a key question in the effort to unify the diverse appearance of AGN.

\iras\ revealed a large population of galaxies whose luminosity was strongly
dominated by the far infrared (Soifer, Houck, and Neugebauer 1987).
(Rieke [1972] had found early indications of a class of ultraluminous
infrared galaxies.)  The infrared emission is thermal emission from dust,
energized in many cases by star formation but in some cases by an AGN.  One
suggested scenario was that some event, possibly a galactic merger, injected
large quantities of gas and dust into the nucleus.  This fueled a luminous
episode of accretion onto a black hole, at first enshrouded by
the dusty gas, whose dissipation revealed the AGN at optical and
and ultraviolet wavelengths (Sanders \etal\ 1988).

\subsection{The Black Hole Paradigm}

The intriguing paper by Lynden-Bell (1969) still did not launch a widespread
effort to understand AGN in terms of accretion disks around black holes. 
Further impetus came from the discovery of black holes of stellar
mass in our Galaxy.  Among the objects discovered by {\em Uhuru} and
other early X-ray experiments were sources involving binary star systems
with a neutron star or black hole.  ``X-ray pulsars'' emitted regular
pulses of X-rays every few seconds as the neutron star turned on its axis. 
The X-ray power was essentially thermal emission from gas transferred from the
companion star, impacting on the neutron star with sufficient velocity to
produce high temperatures.  Another class of source, exemplified by Cyg X-1,
showed no periodic variations but a rapid flickering (Oda \etal\ 1971)
indicating a very small size. 
Analysis of the orbit gave a mass too large to be a neutron star or white
dwarf, and the implication was that the system contained a black hole
(Webster and Murdin 1972; Tananbaum \etal\ 1972).  The X-ray emission was
attributed to gas from the companion O-star heated to very high
temperatures as it spiraled into the black hole by way of a disk (Thorne and
Price 1975).

Galactic X-ray sources, along with cataclysmic variable stars, protostars, and
AGN, stimulated efforts to develop the theory of accretion disks.  In many
cases, the disk was expected to be geometrically thin, and the structure in
the vertical and radial directions could be analyzed separately.  A key
uncertainty was the mechanism by which angular momentum is transported
outward as matter spirals inward.  In a highly influential paper, Shakura and
Sunyaev (1973) analyzed disks in terms of a dimensionless parameter $\alpha$
that characterized the stresses that led to angular momentum transport and
local energy release.  General relativistic corrections were added by
Novikov and Thorne (1973).  This ``$\alpha$-model'' remains the
standard approach to disk theory, and only recently have detailed mechanisms
for dissipation begun to gain favor (Balbus and Hawley 1991).  The
$\alpha$-model gave three radial zones characterized by the relative
importance of radiation pressure, gas pressure, electron scattering, and
absorption opacity.  The power producing regions of AGN disks would fall in
the ``inner'' zone dominated by radiation pressure and electron scattering. 
Electron scattering would dominate in the atmosphere as well as the interior,
and modify the local surface emission from an approximate black body
spectrum. The ``inner'' disk zone suffers both thermal and viscous
instabilities (Pringle 1976; Lightman and Eardley 1974), but the
ultimate consequence of these was unclear.  A model in which the ions and
electrons had different, very high temperatures was proposed for Cyg X-1 by
Eardley, Lightman, and Shapiro (1975).  This led to models of ``ion supported
tori'' for AGN (Rees \etal\ 1982). The related idea 
of ``advection dominated accretion
disks''or ``ADAFs'' (Narayan and Yi 1994) recently has attracted attention.

A key question was, do expected physical processes in disks explain the
phenomena observed in AGN?  In broad terms, this involved producing the
observed continuum and, at least in some objects, generating relativistic
jets, presumably along the rotation axis.  Shields (1978b) proposed that the
flat blue continuum of 3C 273 was thermal emission from the
surface of an accretion disk around a black hole.  For a mass $\sim
10^9~\msun$ and accretion rate $~3~\msunyr$, the size and temperature of the
inner disk was consistent with the observed blue continuum.
This component dominated an assumed nonthermal power law, which would explain
the infrared upturn and the X-rays.  Combining optical, infrared, and
ultraviolet observations, Malkan (1983) successfully fitted the continua of a
number of QSOs with accretion disk models.  Czerny and Elvis (1987)
suggested that the soft X-ray excess of some AGN could be the high frequency
tail of the thermal disk component or ``Big Blue Bump'', which appeared to
dominate the luminosity of some objects.  

Problems confronted the simple picture of thermal emission from a disk
radiating its locally produced energy.  Correlated continuum variations
at different wavelengths in the optical and ultraviolet were
observed in the optical and ultraviolet on timescales shorter than the
expected timescale for viscous or thermal processes to modify the
surface temperature distribution in an AGN disk 
(\eg, Clavel, Wamsteker, and Glass
1989; Courvoisier and Clavel 1991).  This suggested that reprocessing of X-rays
incident on the disk made a substantial contribution to the optical and
ultraviolet continuum (Collin-Souffrin 1991).  Also troublesome
was the low optical polarization observed in normal QSOs, typically one
percent or less.  The polarization generally is oriented parallel to the disk
axis, when this can be inferred from jet structures (Stockman, Angel, and Miley
1979).  Except for face on disks, electron
scattering in disk atmospheres should produce strong polarization oriented
perpendicular to the axis.  Yet another problem was the prediction of strong
Lyman edge absorption features, given effective temperatures similar to those
of O stars (Kolykhalov and Sunyaev 1984).  These issues remain
under investigation today.

The question of fueling a black hole in a galactic nucleus has been
difficult.  Accretion rates of only a few solar masses a year suffice to
power a luminous quasar, and even a billion solar masses is a small
fraction of the mass of a QSO host galaxy.  However, the specific
angular momentum of gas orbiting a black hole at tens or hundreds of
gravitational radii is tiny compared to that of gas moving with normal speeds
even in the central regions of a galaxy.  
The angular momentum must be removed if the gas is to feed the black
hole.  Moreover, some galaxies with massive central black holes are not
currently shining.  Indeed, the rapid increase in the number of quasars with
increasing look back time (Schmidt 1972), implies that there are many
dormant black holes in galactic nuclei.  What caused some to blaze forth as
QSOs while others are inert?  A fascinating possibility was the tidal
disruption of stars orbiting close to the black hole (Hills 1975).  
However, the rate at which new stars would have their orbits evolve into
disruptive ones appeared to be too slow to maintain a QSO luminosity (Frank
and Rees 1976).  The probability  of an AGN in a galaxy appeared to be
enhanced if it was interacting with a nearby galaxy (Adams 1977; Dahari 1984),
which suggested that tidal forces could induce gas to sink into the galactic
nucleus. There,  unknown processes might relieve it of its angular
momentum and allow it to sink closer and closer to the black hole.

The growing acceptance of the black hole model resulted, not from any one
compelling piece of evidence, but rather from the accumulation of observational
and theoretical arguments suggestive of black holes and from the lack of viable
alternatives (Rees 1984).

\subsection{Unified Models}

After the discovery of QSOs, the widely different appearances of different
AGN became appreciated. The question arose, what aspects of this diversity
might result from the observer's location relative to the AGN?  A basic 
division was between radio loud and radio quiet objects.  Since the extended
radio sources radiate fairly isotropically, their presence or absence could
not be attributed to orientation.  Furthermore, radio loud objects seemed
to be associated with elliptical galaxies, and radio quiet AGN with
spiral galaxies.  The huge range of luminosities from Seyferts to QSOs
clearly was largely intrinsic.  However, some aspects could be a
function of orientation.  
 Blandford and
Rees (1978) proposed that BL Lac objects were radio galaxies viewed down the
axis of a relativistic jet.  Relativistic beaming caused the nonthermal
continuum to be very bright when so viewed, and the emission lines (emitted
isotropically) would be weak in comparison.  The same object, viewed from
the side, would have normal emission-line equivalent widths, and the radio
structure would be dominated by the extended lobes rather than the core.

A key breakthrough occurred as a result of advances in the techniques of
spectropolarimetry. Rowan-Robinson (1977) had raised 
the possibility that the BLR of Seyfert 2
galaxies was obscured by dust, rather than being truly absent. 
Using a sensitive spectropolarimeter on the 120-inch Shane telescope at Lick
Observatory, Antonucci and Miller (1985) found that the polarized flux of NGC
1068, the prototype Seyfert 2, had the appearance of a normal Seyfert 1
spectrum.  This was interpreted in terms of a BLR and central continuum
source obscured from direct view by an opaque, dusty torus.  Electron
scattering material above the nucleus near the axis of the torus scattered
the nuclear light to the observer, polarizing it in the process.  This
allowed Seyfert 2's to have a detectable but unreddened continuum. However,
the broad lines had escaped notice because the scattered light was feeble
compared with the narrow lines from the NLR, which was outside the presumed
obscuring torus.  The same object, viewed face on, would be a Seyfert 1. 
Such a picture had also been proposed  by Antonucci (1984) for the broad line
radio galaxy 3C 234. Various forms of toroidal geometry had been anticipated
by Osterbrock (1978) and others, and the idea received support from the
discovery of ``ionization cones'' in the nuclei of some AGN (Pogge 1988). 
Orientation indicators were developed involving the ratio of the core and
extended radio luminosities (Orr and Browne 1982; Wills and Browne 1986). The
concepts of a beamed nonthermal continuum and an obscuring, equatorial torus
remain fundamental to current efforts to unify AGN.
Consideration of the obscuring torus supports the idea that the
X-ray background is produced mostly by AGN (Setti and Woltjer 1989).

\section{THE VIEW FROM HERE}

The efforts described above led to many of the observational and theoretical
underpinnings of our present understanding of AGN.  The enormous
effort devoted to AGN in recent years has led to many further
discoveries and posed exciting challenges.

Massive
international monitoring campaigns (Peterson 1993) have revealed
ionization stratification with respect to radius in the BLR, that the
BLR radius increases with luminosity, and that the gas is not
predominantly in a state of radial flow inwards or outwards.  This
suggests the likelihood of orbiting material.  Models involving a mix of
gas with a wide range of densities and radii may give a natural
explanation of AGN line ratios (Baldwin \etal\ 1995).  Chemical
abundances in QSOs have been analyzed in the context of galactic chemical
evolution (Hamann and Ferland 1993). Recent theoretical work indicates
that the observed, centrally peaked line profiles can be obtained from a
wind leaving the surface of a Keplerian disk (Murray and Chiang 1997).

Efforts to understand the broad absorption lines (BALs) of QSOs have
intensified in recent years.  The geometry and acceleration mechanism
are still unsettled, although disk winds may be involved here too
(Murray \etal\ 1995). Partial coverage of the continuum
source by the absorbing clouds complicates the effort to determine 
chemical abundances (\eg, Arav 1997).

The black hole model has gained support from indirect evidence for massive
black holes in the center of the Milky Way and numerous nearby galaxies
(see Rees 1997).  This includes the remarkable ``H$_2$O megamaser'' VLBI
measurements of the Seyfert galaxy NGC 4258 (Miyoshi \etal\ 1995), which give
strong evidence for a black hole of mass $4\times10^7~\msun$.  
X-ray observations suggest reflection of X-rays incident on an
accretion disk (Pounds \etal\ 1989), and extremely broad
Fe K$\alpha$ emission lines may give a direct look at material
orbiting close to the black hole (Tanaka \etal\ 1998).
These
results reinforce the black hole picture, but much remains to be done to
understand the physical processes at work in AGN.  In spite of
much good work, the origin and fueling of
the hole, the physics of the disk, and the jet production mechanism still are
not well understood.

The nature of the AGN continuum remains unsettled; for example, the
contribution of the disk to the optical and ultraviolet continuum is still
debated (Koratkar and Blaes 1999).  The primary X-ray emission mechanism and
the precise role of thermal and nonthermal emission in the infrared remain
unclear (Wilkes 1999).  Blazars have proved to be strong $\gamma$-ray 
sources, with detections up to TeV energies (Punch \etal\ 1992).

Radio emission was key to the discovery of quasars, and radio techniques have
seen great progress.  The Very Large Array in New Mexico has produced
strikingly detailed maps of radio sources, and shown the narrow channels of
energy from the nucleus to the extended lobes. 
Maps of ``head-tail'' sources in clusters of
galaxies shows the interplay between the active galaxy
and its environment.  The Very
Long Baseline Array (VLBA) will yield improved measurements of structures on
light-year scales in QSOs and provide insights into relativistic motions in
AGN.  Likewise, new orbiting X-ray observatories promise great advances in
sensitivity and spectral resolution.

The Hubble Deep Field and other deep galaxy surveys have led to the
measurement of redshifts for galaxies as high as those of QSOs.  This is
already stimulating increased efforts to understand the interplay between AGN
and the formation and evolution of galaxies. 

The decline of AGN as an active subject of research is nowhere in sight.

\section{BIBLIOGRAPHY}

In addition to the primary literature, I have drawn on a
number of reviews, books, and personal
communications.  For the early work in radio
astronomy, the books by Sullivan (1982,
1984) were informative and enjoyable; the former conveniently
reproduces many of the classic papers.  The book by Burbidge and Burbidge
(1967) was an invaluable guide. A
brief summary of early studies is contained in the introduction to 
Osterbrock's (1989)  book. 
The {\it Conference on Seyfert Galaxies and
Related Objects} (Pacholczyk and Weymann 1968) makes fascinating
reading today. The status of AGN research in the late 1970s is
indicated by the  {\it Pittsburgh Conference on BL Lac
Objects} (Wolfe 1978). Many aspects of AGN are discussed in the
volume in honor of Professor Donald E. Osterbrock (Miller 1985),
which remains of interest both from an historical and a modern
perspective. 

Review articles that especially influenced this work include those by
Bregman (1990) on the continuum; Mushotzky, Done, and Pounds
(1993) and Bradt, Ohashi, and Pounds (1992)
on X-rays; and Stein and Soifer (1983) on dust in galaxies.  Historical
details of the discovery of QSO redshifts are given by Schmidt (1983, 1990);
and an historical account of early AGN studies is given in the
introduction to the volume by Robinson
\etal\ (1964). A comprehensive early review of AGN was given by Burbidge
(1967b).  A review of superluminal radio sources is given by
Kellermann (1985), and the emission-line regions are reviewed by
Osterbrock and Mathews (1986).  A succinct review of important papers in the
history of AGN research is given by Trimble (1992).

Recent books on AGN include those of
Krolik (1999), Peterson (1997), and Robson (1996).
Many interesting articles are contained
in the volume edited by Arav \etal\ (1997).  
Recent technical reviews include
those by Koratkar and Blaes (1999) on the disk continuum; Antonucci (1993)
and Urry and Padovani (1995) on unified models; Lauroesch \etal\ (1996) on
absorption lines and chemical evolution; Ulrich, Maraschi, and Urry (1997)
on variability; and Hewett
and Foltz (1994) on quasar surveys.

\acknowledgments

The author is indebted to many colleagues
for valuable
communications and comments on the
manuscript, including Stu Bowyer, 
Geoff and Margaret Burbidge, Marshall Cohen, Suzy Collin, Martin
Elvis, Jesse Greenstein, Ken Kellermann, Matt Malkan, 
Bill Mathews, Richard Mushotzky, Gerry
Neugebauer, Bev Oke, Martin Rees, George Rieke,
Maarten Schmidt, Woody Sullivan, Marie-Helene Ulrich, and
Bev and Derek Wills.  
Don Osterbrock was especially
supportive and helpful.   
This article was written in part during visits to the Department of
Space Physics and Astronomy,
Rice University; Lick Observatory; and the Institute for
Theoretical Physics, University of California, Santa
Barbara.  The hospitality of these institutions is
gratefully acknowledged.  
This work was supported in part by The Texas Higher
Education Coordinating Board.

\def\astrofiz{ Astrofizica}
\def\procnas{ Proc.~Nat.~Acad.~Sci.}
\def\ausjsr{ Austral.~J.~Sci.~Res.}
\def\procire{ Proc.~IRE}
\def\ausjphys{ Austral.~J.~Phys.}
\def\pr{ Phys.~Rev.}
\def\prl{ Phys.~Rev.~Lett.}
\def\dokl{ Dokl.~Akad.~Nauk~SSSR}
\def\lowell{ Lowell~Obs.~Bull.}
\def\observ{ Observ.}
\def\science{ Science}

\end{document}